\documentclass[11pt]{article}
\usepackage{amssymb}
\usepackage{graphics}
\usepackage{epsfig}
\usepackage{a4wide}
\usepackage{slashed}
\usepackage{CJK}

\begin{document}

\newcommand{\nn}{\nonumber}
\newcommand{\ppJW}{$pp \to J/\psi +W+X~$}

\title{  $J/\psi$ production associated with a $W$-boson at the $7~TeV$ LHC  }
\author{ Song Mao$^a$, Li Gang$^a$, Ma Wen-Gan$^b$, Zhang Ren-You$^b$, Guo Lei$^b$, and Guo Jian-You$^a$  \\
{\small  $^a$ School of Physics and Material Science, Anhui University, Hefei, Anhui 230039, P.R.China}  \\
{\small  $^b$ Department of Modern Physics, University of Science and Technology of China (USTC),}  \\
{\small   Hefei, Anhui 230026, P.R.China}}


\date{}
\maketitle \vskip 15mm
\begin{abstract}
We calculate the complete next-to-leading order (NLO) QCD corrections
to the $J/\psi +W$ associated production within the factorization formalism
of nonrelativistic QCD at the $7~TeV$ LHC. We provide the
numerical results for the leading-order (LO), NLO QCD corrected
differential cross sections of the $J/\psi$ transverse momentum by
adopting the event selection criteria requested by the ATLAS experiment.
We find that the differential cross section at the LO is
significantly enhanced by the NLO QCD corrections.
\end{abstract}


\vfill \eject \baselineskip=0.32in

\renewcommand{\theequation}{\arabic{section}.\arabic{equation}}
\renewcommand{\thesection}{\Roman{section}.}
\newcommand{\nb}{\nonumber}

\newcommand{\Dir}{\kern -6.4pt\Big{/}}
\newcommand{\Dirin}{\kern -10.4pt\Big{/}\kern 4.4pt}
\newcommand{\DDir}{\kern -7.6pt\Big{/}}
\newcommand{\DGir}{\kern -6.0pt\Big{/}}

\makeatletter      
\@addtoreset{equation}{section}
\makeatother       

\par
Recently, after we published our work on the QCD corrections to
$J/\psi$ production in association with a $W$-boson at the $14~TeV$
LHC \cite{liw}, Darren Price in the ATLAS experiment suggested us to
extend our study to the colliding energy of $7~TeV$, and asked for
theoretical predictions for the corresponding differential cross
sections in the restricted kinematic regions of the final particles.
In Ref.\cite{liw} we presented the detailed calculations of the NLO
QCD corrections to the $J/\psi +W$ associated production at the
$\sqrt{s}=14~TeV$ LHC, but not for the $7~TeV$ LHC. In this letter,
we calculate the NLO QCD corrections to the associated $J/\psi$
production with a $W$ gauge boson in the nonrelativistic QCD (NRQCD)
at the $7~TeV$ LHC, and provide the theoretical predictions for the
$p_T$ distribution of $J/\psi$ by adopting the event selection
criteria requested by the ATLAS experiment. Previous theoretical studies \cite{liw,kniehl02} suggest
$W^{\pm}$ +prompt $J/\psi$ production should be dominated by colour octet processes. However, recent work \cite{Lansberg} suggests
that in 7 TeV $pp$ collisions color singlet (CS) and color octet (CO) contributions are comparable.
Measurements of the production cross sections can help to distinguish between these models.

\par
\begin{figure}
\begin{center}
\includegraphics[scale=0.65]{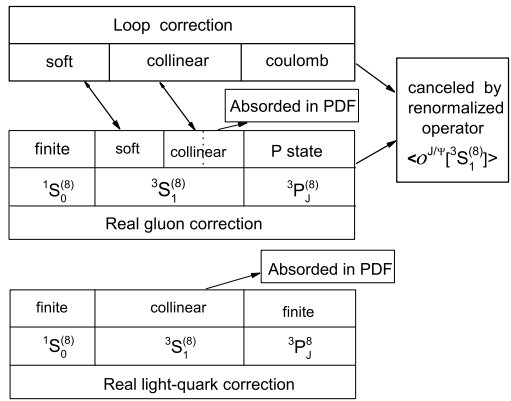}
\vspace*{-0.3cm} \caption{\label{fig2}  The divergence structure and
divergence cancellation routes in the NLO calculations for the $pp
\to J/\psi+W^++X$ process.}
\end{center}
\end{figure}

We calculate the \ppJW process by applying the covariant-projector
method \cite{projectors} in the NRQCD framework.
In the leading order (LO) calculations, only the Fock state
$^3S_1^{(8)}$ is involved. However, for the NLO QCD corrections to
this process, we need to consider both the virtual corrections and
the real gluon/light-quark emission processes. The virtual
corrections only contain the contribution from $c\bar{c}$ Fock state
$^3S_1^{(8)}$, while the real gluon/light-quark emission processes
should involve $^1S_0^{(8)}$, $^3S_1^{(8)}$ and $^3P_J^{(8)}$ Fock
states. In our calculations, we apply the dimensional regularization
(DR) scheme to regularize the UV and IR divergences, and the
modified minimal subtraction ($\overline{{\rm MS}}$) and
on-mass-shell schemes to renormalize the strong coupling constant
and the quark wave functions, respectively. In Fig.\ref{fig2}, we
present the divergence structure and divergence cancellation routes
in the NLO calculation for the $pp \to J/\psi+W^++X$ process.
The two cutoff phase space slicing method
(TCPSS)\cite{TCPSS} has been employed for dealing with the soft and
collinear singularities in real gluon/light-quark emission
corrections.

\par
We take CTEQ6L1 PDFs with the one-loop running $\alpha_s$ in the LO
calculations and CTEQ6M PDFs with the two-loop $\alpha_s$ in the NLO
calculations \cite{CTEQ6}, respectively. For simplicity we define
$\mu \equiv \mu_r=\mu_f$ and take the renormalization/factorization
and NRQCD scales as $\mu= m_T^{J/\psi}$ or $m_W$, $\mu_{\Lambda} =
m_c$, respectively, where $m_T^{J/\psi} = \sqrt{\Big( p_T^{J/\psi}
\Big)^2 + m_{J/\psi}^2}$ is the $J/\psi$ transverse mass. The masses
of the external particles and the fine structure constant are taken
as $m_W =80.398~{\rm GeV}$, $m_c = \frac{1}{2}m_{J/\psi} = 1.5~{\rm
GeV}$ and $\alpha = 1/137.036$. We use two sets of color-octet (CO)
long distance matrix elements (LDME) listed in Table \ref{tab:set}
as the input parameters, which were obtained by fitting the
differential cross section and polarization of inclusive $J/\psi$
production data from various experiments \cite{kniehl1201,chao1201}.
\begin{table}[t]
  \centering
  \begin{tabular}{|c|c|c|}
  \hline\hline
          & Set 1 ($GeV^3$)& Set 2 ($GeV^3$) \\
\hline
  $<{\cal O}^{J/\psi}[^1S_0^{(8)}]>$ & $3.04 \times 10^{-2}$ & $8.9 \times 10^{-2}$ \\
  $<{\cal O}^{J/\psi}[^3S_1^{(8)}]>$ & $1.68 \times 10^{-3}$ & $3 \times 10^{-3}$ \\
  $<{\cal O}^{J/\psi}[^3P_0^{(8)}]>$ & $-9.08 \times 10^{-3}$ &$5.6 \times 10^{-3}$ \\

\hline\hline
\end{tabular}
\caption{Two sets of CO LDME for $J/\psi$. Set 1 are extracted via a
global fit from various hadroproduction, photoproduction, two-photon
scattering and electron-positron annihilation experiments
\cite{kniehl1201}. Set 2 are extracted by fitting the differential
cross section and polarization of prompt $J/\psi$ simultaneously at
the Tevatron \cite{chao1201}.}
 \label{tab:set}
\end{table}

\par
We calculate the LO and NLO QCD corrected cross sections for the $pp
\to J/\psi+W \to J/\psi+\mu+\nu_{\mu}+X$ process at the $7~TeV$ LHC
by adopting the following kinematic cuts on the final particles
according to the requests of the ATLAS data analyses:
\begin{equation}\label{cuts}
p_{T}^{\mu} > 25 {~GeV}, ~~|\eta^{\mu}| < 2.4, ~~
p_{T}^{\nu} > 20 {~GeV}, ~~ m_T^W > 40 {~GeV}.
\end{equation}
where $m_T^W$ is $W$ boson transverse mass defined as $m_T^W =
\sqrt{2 p^\mu_T p^{\nu}_T\left[1-\cos\left(\phi^{\mu}-\phi^{\nu}\right)\right]}$, and
$\phi^{\mu}-\phi^{\nu}$ is the angle between muon and
neutrino in the transverse plane \cite{Atlas}.

\par
In our calculations the uncertainty for theoretical prediction
mainly comes from two parts: the short distance part and the long
distance part. For the former part, the dependence of the cross
section on the renormalization scale $\mu_r$ and factorization scale
$\mu_f$ induces theoretical uncertainty. For the later part, the
values of CO LDME for the $J/\psi$ extracted from the experiments by
different experimental groups vary significantly. In the following
tables and figures, we will give the results by adopting two set of
CO LDME and two typical scales ($\mu= m_T^{J/\psi},~m_W$).
\begin{table}[t]
  \centering
  \begin{tabular}{c|c|c|c|c|c|c|c|c}
  \hline\hline
 $p_T^{J/\psi} $ [GeV]& \multicolumn{4}{|c|}{$\mu = m_T^{J/\psi}$ }& \multicolumn{4}{|c}{$\mu = M_W$ } \\
\hline & \multicolumn{2}{|c|}{$|y_{J/\psi} |<1.0$ }&
\multicolumn{2}{|c|}{$1.0<|y_{J/\psi}|<2.1$ }
& \multicolumn{2}{|c|}{ $|y_{J/\psi} |<1.0$}&\multicolumn{2}{|c}{$1.0<|y_{J/\psi}|<2.1$} \\
\hline
&  LO [fb] & NLO [fb] & LO [fb]& NLO [fb]  & LO [fb] & NLO [fb] & LO [fb] & NLO [fb] \\
\hline
 8.5$\sim$10 & 0.567 & 2.96 & 0.531 & 2.87 & 0.361 & 1.75  & 0.336  & 1.45  \\
 10$\sim$14 & 0.960 & 4.86 & 0.899 & 4.65 & 0.637  & 3.10  & 0.593  & 2.52 \\
 14$\sim$18 & 0.527 & 2.56 & 0.494 & 2.42 & 0.369  & 1.81  & 0.344  & 1.45\\
 18$\sim$22 & 0.315 & 1.51 & 0.296 & 1.40 & 0.230  & 1.14  & 0.215  & 0.896 \\
 22$\sim$30 & 0.331 & 1.54 & 0.311 & 1.44 & 0.253  & 1.27  & 0.238  & 0.990 \\
\hline\hline
\end{tabular}
\caption{The cross sections for the $pp \to J/\psi+W \to
J/\psi+\mu+\nu_{\mu}+X$ process at the $7~TeV$ LHC in the
$p_T^{J/\psi}$ and the $J/\psi$ rapidity intervals by taking the Set
1 LDME parameters.} \label{tab:ptcross1}
\end{table}

\begin{table}[t]
  \centering
  \begin{tabular}{c|c|c|c|c|c|c|c|c}
  \hline\hline
 $p_T^{J/\psi} $ [GeV]& \multicolumn{4}{|c|}{$\mu = m_T^{J/\psi}$ }& \multicolumn{4}{|c}{$\mu = M_W$ } \\
\hline & \multicolumn{2}{|c|}{$|y_{J/\psi} |<1.0$ }&
\multicolumn{2}{|c|}{$1.0<|y_{J/\psi}|<2.1$ }
& \multicolumn{2}{|c|}{ $|y_{J/\psi} |<1.0$}&\multicolumn{2}{|c}{$1.0<|y_{J/\psi}|<2.1$} \\
\hline
&  LO [fb] & NLO [fb] & LO [fb]& NLO [fb]  & LO [fb] & NLO [fb] & LO [fb] & NLO [fb] \\
\hline
 8.5$\sim$10 & 1.01 & 3.51 & 0.949 & 3.66 & 0.645 & 2.58  & 0.600  & 1.91  \\
 10$\sim$14 & 1.71 & 5.64 & 1.60 & 5.69 & 1.14  & 4.54  & 1.06  & 3.19 \\
 14$\sim$18 & 0.942 & 2.91 & 0.883 & 2.82 & 0.659  & 2.61  & 0.615  & 1.77\\
 18$\sim$22 & 0.562 & 1.67 & 0.528 & 1.59 & 0.410  & 1.62  & 0.384  & 1.06 \\
 22$\sim$30 & 0.591 & 1.67 & 0.555 & 1.57 & 0.452  & 1.80  & 0.424  & 1.13 \\
\hline\hline
\end{tabular}
\caption{The cross sections for the $pp \to J/\psi+W \to
J/\psi+\mu+\nu_{\mu}+X$ process at the $7~TeV$ LHC in the
$p_T^{J/\psi}$ and the $J/\psi$ rapidity intervals by taking the Set
2 LDME parameters.} \label{tab:ptcross2}
\end{table}

\par
In Table \ref{tab:ptcross1} and Table \ref{tab:ptcross2}, we provide
the cross sections for the $pp \to J/\psi+W \to
J/\psi+\mu+\nu_{\mu}+X$ process at the $7~TeV$ LHC in the
$p_T^{J/\psi}$ and the $J/\psi$ rapidity intervals by taking the two
different set of LDME parameters. In each table, we list the
differential cross sections for $\mu_r = \mu_f = m_T^{J/\psi}$ and
$\mu_r = \mu_f =m_W$, respectively. We adopt the event select
criteria with the cuts of $p_{T}^{\mu} > 25 {~GeV}$, $|\eta^{\mu}| <
2.4$, $p_{T}^{\rm miss} > 20 {~GeV}$ and $m_T^W > 40 {~GeV}$ as
requested by ATLAS experiment where muon and neutrino are the $W$
decay products. In these two tables we list the $J/\psi$ production
rates in the following $p_T^{J/\psi}$ intervals of $8.5 \sim 10
~GeV$, $10 \sim 14 ~GeV$, $14 \sim 18 ~GeV$, $18 \sim 22 ~GeV$, $22
\sim 30 ~GeV$, and the $J/\psi$ rapidity intervals of
$|y_{J/\psi}|<1.0$ and $1.0<|y_{J/\psi}|<2.1$, separately. The
cross section for the $pp \to J/\psi+W \to J/\psi+\mu+\nu_{\mu}+X$
process is obtained via the cross section for $pp \to J/\psi+W+X$
multiplying the branching fraction for the $W \to \mu \nu_{\mu}$
decay, which is taken as $10.57 \% $ \cite{pdg2012}.

\begin{figure}
\includegraphics[scale=0.75]{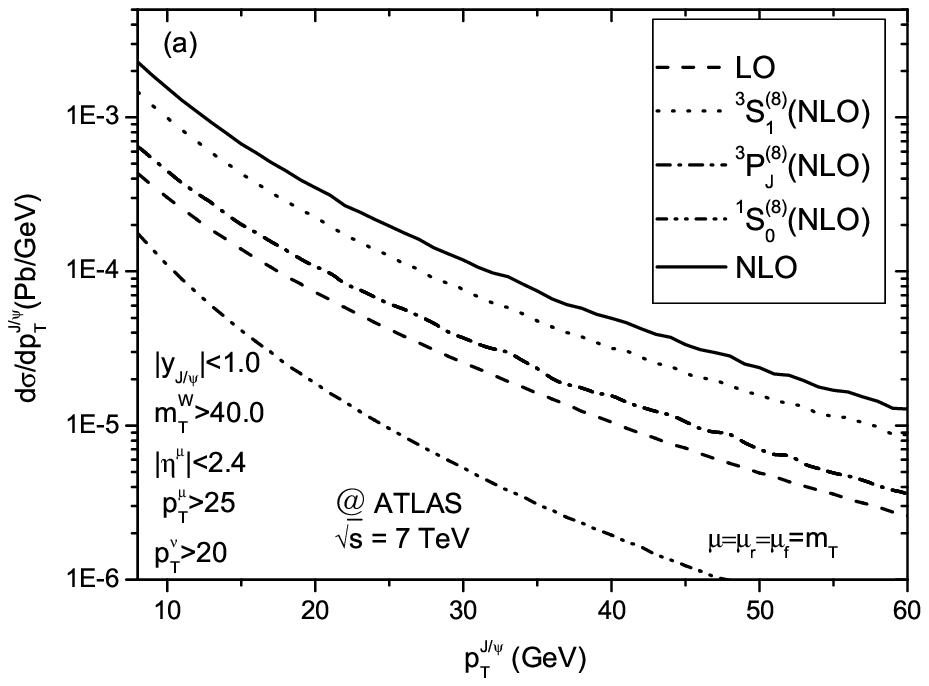}
\includegraphics[scale=0.75]{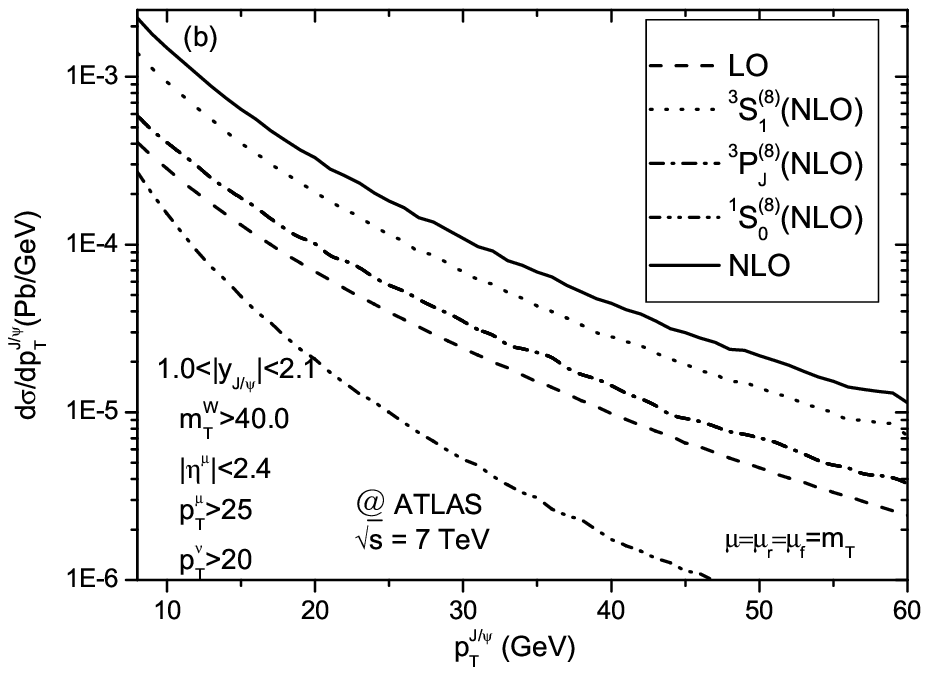}
\vspace*{-0.3cm} \caption{\label{fig3}  The LO and NLO QCD corrected
distributions of the transverse momentum of $J/\psi$ for the $pp \to
J/\psi+W \to J/\psi+\mu+\nu_{\mu}+X$ process at the $7~TeV$ LHC in
the region with the constraints on final particles shown in
(\ref{cuts}) by taking the Set 1 LDME parameters and $ \mu= \mu_r =
\mu_f = m_T^{J/\psi}$. (a) $|y_{J/\psi}|<1.0$, (b)
$1.0<|y_{J/\psi}|<2.1$. }
\end{figure}

\begin{figure}
\includegraphics[scale=0.75]{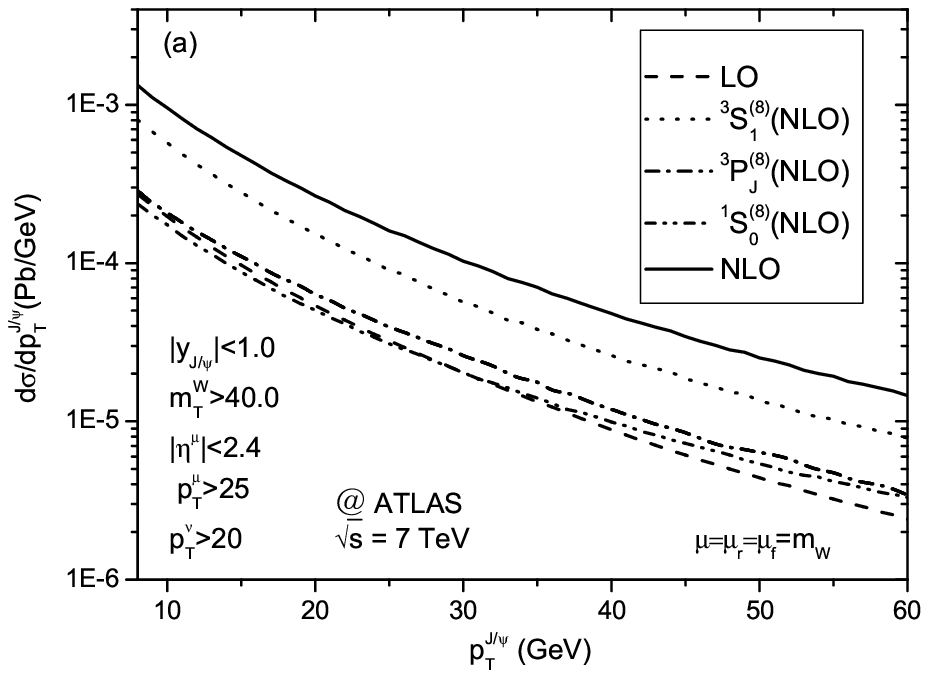}
\includegraphics[scale=0.75]{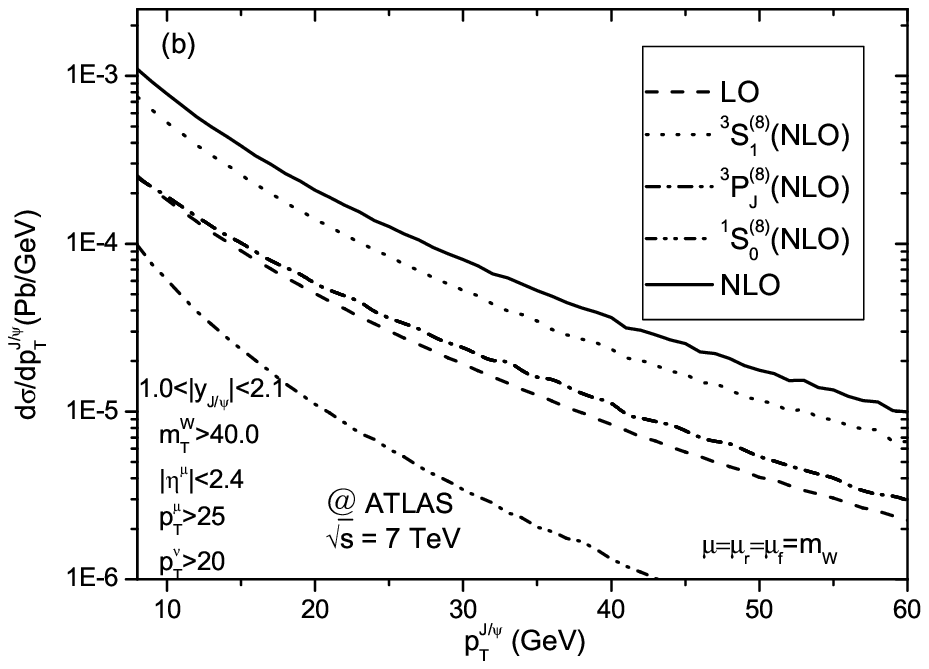}
\vspace*{-0.3cm} \caption{\label{fig4} The LO and NLO QCD corrected
distributions of the transverse momentum of $J/\psi$ for the $pp \to
J/\psi+W \to J/\psi+\mu+\nu_{\mu}+X$ process at the $7~TeV$ LHC in
the region with the constraints on final particles shown in
(\ref{cuts}) by taking the Set 1 LDME parameters and $ \mu= \mu_r =
\mu_f = m_W$. (a) $|y_{J/\psi}|<1.0$, (b) $1.0<|y_{J/\psi}|<2.1$.}
\end{figure}

\par
In Figs.\ref{fig3}(a,b) and Figs.\ref{fig4}(a,b), we present the the
LO and NLO QCD corrected distributions of the transverse momentum of
$J/\psi$ for the $pp \to J/\psi+W \to J/\psi+\mu+\nu_{\mu}+X$
process at the $7~TeV$ LHC in the region restricted by the
conditions shown in (\ref{cuts}) by adopting the Set 1 LDME
parameters. The renormalization/factorization scales are set to be
$\mu=m_T^{J/\psi}$ and $\mu=m_W$ in Figs.\ref{fig3} and
Figs.\ref{fig4}, separately. In Figs.\ref{fig3}(a,b) and
Figs.\ref{fig4}(a,b), we give the predictions for the transverse
momentum distributions of $J/\psi$ in two rapidity intervals on the
$J/\psi$: $|y_{J/\psi}|<1.0$ and $1.0<|y_{J/\psi}|<2.1$. As a
comparison, we also depict the contributions from the $^1S_0^{(8)}$
and $^3S_1^{(8)}$ and $^3P_J^{(8)}$ Fock states in
Figs.\ref{fig3}(a,b) and Figs.\ref{fig4}(a,b). The short distance
part of the differential cross section contributed by the
$^3P_J^{(8)} (J = 0, 1, 2)$ Fock states is negative, and the LDME
$<{\cal O}^{J/\psi}[^3P_0^{(8)}]>$ with Set 1 LDME parameters is
also negative. The differential cross section contributed by the $^3P_J^{(8)} (J = 0,
1, 2)$ Fock states should be positive. These figures show clearly
that the differential cross section at the LO is significantly
enhanced by the QCD corrections.

\begin{figure}
\includegraphics[scale=0.75]{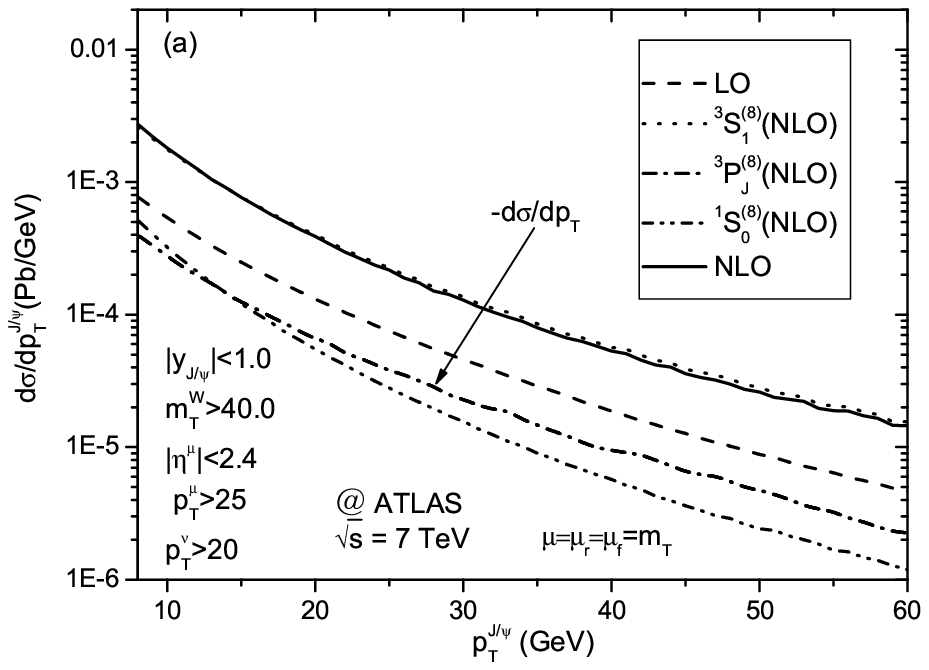}
\includegraphics[scale=0.75]{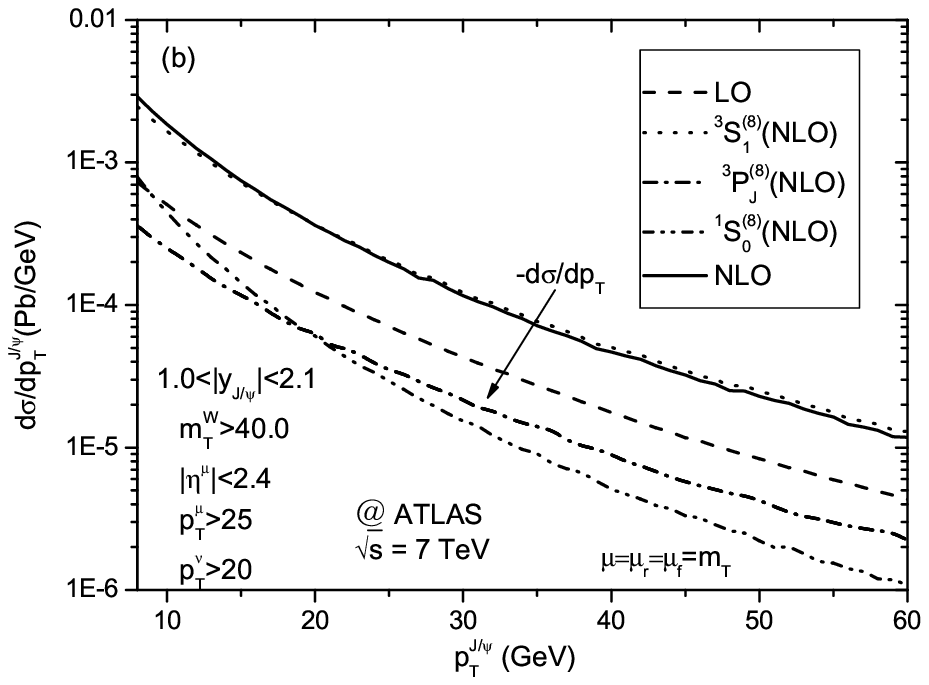}
\vspace*{-0.3cm} \caption{\label{fig5}  The LO and NLO QCD corrected
distributions of the transverse momentum of $J/\psi$ for the $pp \to
J/\psi+W \to J/\psi+\mu+\nu_{\mu}+X$ process at the $7~TeV$ LHC in
the region with the constraints on final particles shown in
(\ref{cuts}) by taking the Set 2 LDME parameters and $ \mu= \mu_r =
\mu_f =  m_T^{J/\psi}$. (a) $|y_{J/\psi}|<1.0$, (b)
$1.0<|y_{J/\psi}|<2.1$.}
\end{figure}

\begin{figure}
\includegraphics[scale=0.75]{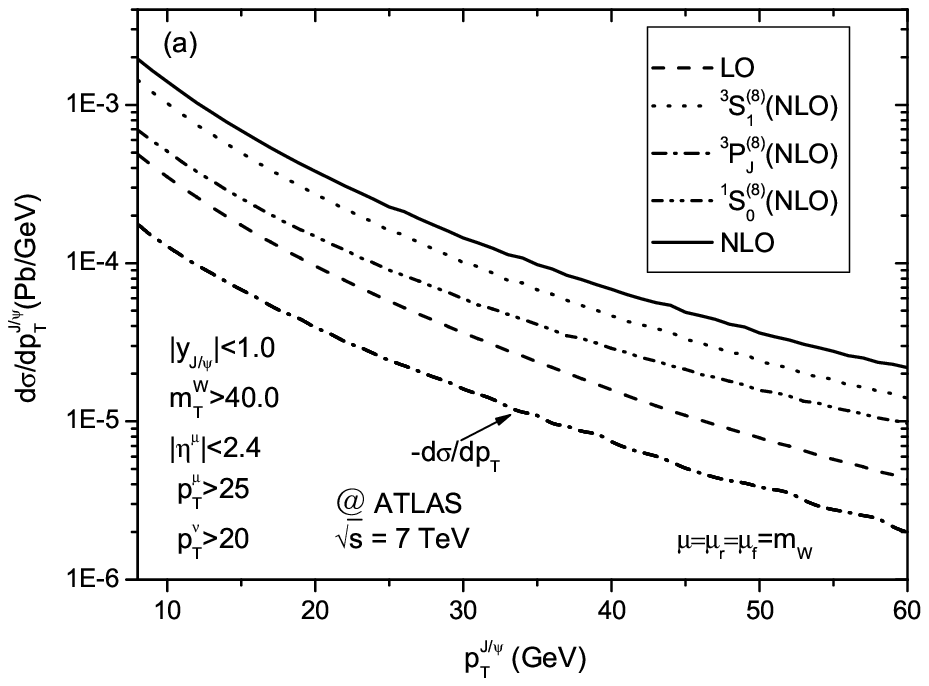}
\includegraphics[scale=0.75]{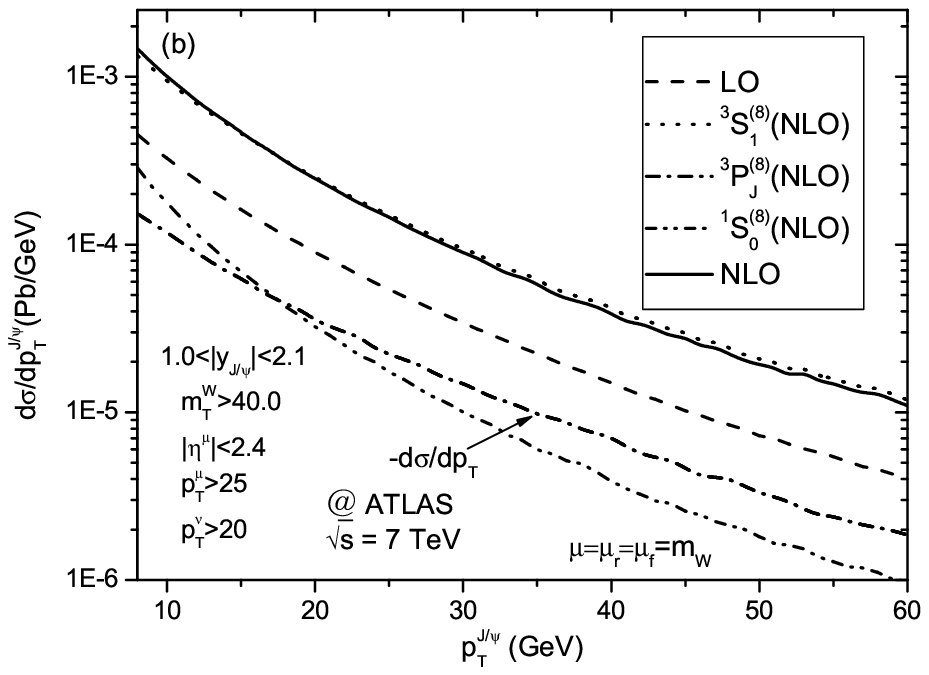}
\vspace*{-0.3cm} \caption{\label{fig6}  The LO, NLO QCD corrected
distributions of the transverse momentum of $J/\psi$ for the $pp \to
J/\psi+W \to J/\psi+\mu+\nu_{\mu}+X$ process at the $7~TeV$ LHC in
the region with the constraints on final particles shown in
(\ref{cuts}) by taking the Set 2 LDME parameters and $ \mu= \mu_r =
\mu_f = m_W$. (a) $|y_{J/\psi}|<1.0$, (b) $1.0<|y_{J/\psi}|<2.1$.}
\end{figure}

\par
In Figs.\ref{fig5}(a,b) and Figs.\ref{fig6}(a,b), we give the LO and
NLO QCD corrected distributions of the transverse momentum of
$J/\psi$ for the $pp \to J/\psi+W \to J/\psi+\mu+\nu_{\mu}+X$
process at the $7~TeV$ LHC taking the Set 2 LDME parameters in the
region restricted by the fiducial cuts on final particles as shown
in (\ref{cuts}). The renormalization/factorization scales are set to
be $ \mu= m_T^{J/\psi}$ and $\mu= m_W$ in Figs.\ref{fig5}(a,b) and
Figs.\ref{fig6}(a,b), separately. We also give the predictions for
the transverse momentum distributions of $J/\psi$ in two $J/\psi$
rapidity intervals: $|y_{J/\psi}|<1.0$ and $1.0<|y_{J/\psi}|<2.1$.
Fig.\ref{fig5}(a), Fig.\ref{fig6}(a) are for $|y_{J/\psi}|<1.0$, and
Fig.\ref{fig5}(b), Fig.\ref{fig6}(b) for $1.0<|y_{J/\psi}|<2.1$. The
short distance contribution part from the $^3P_J^{(8)} (J = 0, 1,
2)$ Fock states is negative, while the LDME $<{\cal
O}^{J/\psi}[^3P_0^{(8)}]>$ with Set 2 LDME parameters is also
positive, thus the contribution from $^3P_J^{(8)} (J = 0, 1, 2)$
Fock states is negative. The negative contributions of $^3P_J^{(8)}
(J = 0, 1, 2)$ Fock states are also presented in
Figs.\ref{fig5}(a,b) and Figs.\ref{fig6}(a,b) for comparison.

\par
In summary, we provide the predictions on the $J/\psi +W$
associated production within the factorization formalism of the
nonrelativistic QCD at the $7~TeV$ LHC using the parameters
and kinematic constraints suggested by the ATLAS experiment. We show
the distributions of the transverse momentum of $J/\psi$ for the $pp
\to J/\psi+W \to J/\psi+\mu+\nu_{\mu}+X$ process in a region constrained
by putting the cuts on final particles shown in (\ref{cuts}), and using the
two set of LDME parameters and two typical energy scales separately.
Our results show that the theoretical uncertainties come from the
choices of the scales and the CO LDME parameters.

\vskip 5mm
\par
\noindent{\large\bf Acknowledgments:} The authors would like thank Mr.
Darren Price for helpful discussion. This work was supported in
part by the National Natural Science Foundation of China
(No.11205003, No.11275190, No.11075150, No.11005101, No. 11175001), the Key
Research Foundation of Education Ministry of Anhui Province of China
(No. KJ2012A021), the Youth Foundation of Anhui Province(No. 1308085QA07),
and financed by the 211 Project of Anhui
University (No.02303319).


\vskip 5mm
\begin{thebibliography}{99}

\bibitem{liw} Li G, Song M, Zhang R Y, and Ma W G,
Phys.\ Rev.\ {\bf D 83} (2011) 014001, arXiv:1012.3798.

\bibitem{kniehl02} Kniehl B A, Palisoc C P, and Zwirner L, Phys. Rev. {\bf D 66} (2002) 114002,
arXiv:hep-ph/0208104.

\bibitem{Lansberg} Lansberg J P and Lorce C, (2013), arXiv:1303.5327.

\bibitem{projectors} Petrelli A, Cacciari M, Greco M, Maltoni F, Mangano M L, Nucl. Phys. {\bf
B 514} (1998) 245.

\bibitem{TCPSS} Harris B W and Owens J F,
 Phys.\ Rev.\ {\bf D 65}, 094032 (2002).

\bibitem{CTEQ6} Pumplin J, Stump D R, Huston J, Lai H L, Nadolsky P, and Tung W K,
JHEP {\bf 0207}, 012 (2002).

\bibitem{kniehl1201} Butenschoen M, Kniehl B A, Nucl.\ Phys.\ {\bf B} Proceedings Supplement XX (2012) 1-11.

\bibitem{chao1201} Chao K T, Ma Y Q, Shao H S, Wang K, Zhang Y J, Phys.\ Rev.\ Lett. {\bf 108}, 242004 (2012).

\bibitem{Atlas} The ATLAS Collaboration, JHEP {\bf 06} (2013) 084, arXiv:1302.2929.

\bibitem{pdg2012} Beringer J set al. (Particle Data Group), Phys. Rev. {\bf D 86}, 010001(2012).

\end{thebibliography}
\end{document}